# Field tests on a full-scale steel chimney subjected to vortex-induced vibrations

Øyvind Mortveit Ellingsen[1,2], Olivier Flamand[1], Xavier Amandolese[2,3], Francois Coiffet[4], Pascal Hémon[2]

[1]CSTB, Nantes, France; [2]LadHyx, CNRS-Ecole polytechnique, Palaiseau, France; [3]LMSSC, CNAM, Paris, France; [4]CERIC, Poujoulat Group, Granzay-Gript, France

## 1  Abstract

Industrial chimneys, launch vehicles and stacks are examples of large diameter circular cross section structures which can be prone to cross-wind vortex-induced vibrations. VIV has been extensively studied for both fundamental and applied issues, but few documented studies concern high Reynolds number regime (> $5 \cdot 10^5$) in atmospheric turbulent wind. This paper introduces a field test on a slender light and low damped chimney designed to experience "supercritical" VIV at moderate wind velocity. The chimney was recently erected in a wind monitored field, near the Atlantic coast of France. The purpose of this paper is to present the first vibration results obtained during a sequential 13-days period in September 2020. A statistical analysis has been performed on the amplitude and dominant frequency responses and results are reported in term of probability distribution as a function of wind speed and direction. VIV events of low (< 15 % of diameter) to moderate amplitude (> 30 % of diameter) have been highlighted in a range of wind velocity 25 % lower than expected, along with significant influence of the wind direction. Low turbulent easterly wind giving vortex-induced vibrations with the highest amplitude.

## 2  Introduction

Slender structures with circular cross section can be prone to vibrations under wind effects and must be designed and/or treated accordingly. For an isolated tower, stack or chimney, one generally considers two kinds of vibrations: in-line vibrations due to atmospheric turbulence and cross-wind vibrations due to the vortex signature. The former concerns extreme wind speed and is the consequence of random aerodynamic load due to turbulence. Since this load can be considered as independent from the structure's dynamic response, the problem can be addressed using well-accepted random vibration methods or simplified equivalent static formulation (see for example [9, 26]). The latter is more complex. It is the consequence of a nonlinear coupling between the fluid force due to the Karman vortex wake and the chimney's motion. This phenomenon, known as vortex-induced vibration (VIV), has been extensively studied for both fundamental and applied issues (see for example [5, 20, 26] for a review).

VIV is characterized by significant oscillations of self-limited amplitude in a limited velocity range where the wake frequency is controlled by the motion, a phenomenon referred as lock-in. Both the oscillation amplitude and range of lock-in strongly depend on the structure to fluid mass ratio and on the damping ratio of the structure. This is encapsulated in the Scruton number (Sc, a dimensionless mass-damping ratio parameter) with low values leading to higher vibration amplitudes and a wider lock-in range. The Reynolds number and turbulence characteristic of the incoming flow can also have significant impact on the VIV of slender structure in atmospheric boundary layer [23, 28]. It is well established that the VIV response is strong for low turbulence flow in the sub-critical Reynolds number regime (< $3 \cdot 10^5$), comparatively negligible in the critical-transitional Reynolds number regime and that a recovering VIV response can be observed in the super-critical Reynolds number regime (> $10^6$) [23].

Tall industrial chimneys, launch vehicles and stacks are examples of large diameter circular cross section structures which can be prone to vortex-induced vibrations at supercritical Reynolds number in atmospheric turbulent wind. It is then necessary to validate appropriate methodology for their design and/or the design of additional damping devices (cf. [4] for simulations on the effect of additional damping devices). VIV models are numerous and codified methods can be found in many standards but their ability to capture the VIV amplitude response at super-critical Reynolds numbers and with real atmospheric boundary layers needs validation. Wind tunnels are also important in response prediction though the scaled models can cause cross-wind loads to be highly overestimated [29]. A method for overcoming the scaling effect is to artificially increase the Reynolds number by adding surface roughness to the cylinder's surface [20, 21]. While this can satisfactorily change the vortex wake signature (e.g. mean drag and rms lift) for a fixed cylinder, the impact on an 3D slender cylinder during lock-in is less clear and wind tunnel studies shows various conclusions so that it needs to be further investigated [2, 6, 11, 27].

Continuous measurements from monitored industrial chimneys can be found in the literature [7, 13, 14, 18, 19, 25, 30] but the monitoring is often limited to acceleration data and a reference velocity. These chimneys have been designed to or treated to limit VIV meaning that the observed vibrations were small. Additionally, the access and opportunity to install extra sensors were limited as they are in use. More response data on industrial chimneys are available but often only the maximum amplitudes are mentioned and are used to validate VIV models in design standards [15, 16].

For other more well-studied circular structures using field-experiments, there is a problem of dimensions [12, 23, 32]. Due to their smaller size, the high amplitude VIV response was at sub-critical or critical Reynolds numbers rather than super-critical. The same problem was observed in wind tunnels when using larger scale wind tunnel experiments [3] as the speed needed to reach super-critical wind speed is large.

In that context, a custom-made 35,5 m steel chimney have been recently erected and instrumented on a monitored wind field, in Bouin (near the Atlantic coast of France). This chimney was designed to have a low Scruton number (Sc < 2) and to experience "super-critical" VIV at moderate wind speeds (< 10 m/s). The paper is organized as follows: field test information and methodology including the structural characteristics of the chimney are presented in Section 3. Characterization of the incoming wind is reported in Section 4. Vibration results obtained during a sequential 13-days period in September 2020 are presented in section 5, before the conclusion and outlooks of this new test platform.

## 3 Field-test platform details and methodology

### 3.1 Structural characteristics of the chimney

The chimney was designed, manufactured and erected by Beirens (Poujoulat group) during the summer of 2020. Figure 1 shows a view of the chimney in the field. This custom-made steel chimney of height h = 35,5 m has a diameter $d_{lower}$ = 1 m for its 12 m long bottom part and a diameter $d_{top}$ = 2 m for its 20,5 m long upper part, with a 3 m long tapered connecting element (see figure 2). This unusual shape (for a chimney) was chosen to ensure vortex-induced vibrations in the super-critical Reynolds number range (> $10^6$) at moderate wind speed (< 10 m/s) for the purpose of this experiment.

Structural characteristics of the chimney are given table 1, along with the expected Reynolds number at the critical wind speed, referred to as $Re_{VIV}$.

*Table 1 Structural characteristic of the chimney (as identified at the start of the experiment).*

| $d_{top}$ [m] | $d_{lower}$ [m] | h [m] | $h_{d=2m}$ [m] | $m_e$ [kg/m] | $f_1$ [Hz] | $\zeta_1$ [%] | Sc [-] | $Re_{VIV}$ [-] |
|---|---|---|---|---|---|---|---|---|
| 2 | 1 | 35,5 | 20,5 | 322,6 | 0,78 | 0,22 | 1,82 | 1,16·$10^6$ |

The equivalent mass, $m_e$, has been calculated using equation (1) where $\psi(z)$ is the shape of the first bending mode and $m(z)$ is the mass per unit height (both being identified during the design phase of the chimney using CAD).

Experimental tests performed at the beginning of the test campaign identified first natural frequency, $f_1$, as 0,78 Hz and allowed the identification of the associated damping ratio $\zeta_1$. The Scruton number, calculated using formula (2), was rather low $Sc$=1,82 and high amplitude response was expected. Using the Eurocode's recommendations [10], the critical wind speed ($f_s$=$f_n$ with a Strouhal number St=0,18) was close to 8,7 m/s ($Re_{VIV} \approx 1{,}16 \cdot 10^6$) and the maximum dimensionless amplitude (in top diameter) would be either 0,31 (method 1 of Ruscheweyh [23]) or 0,53 (method 2 of Vickery and Basu [28]).

$$m_e = \frac{\int_0^h \psi(z)^2 m(z) dz}{\int_0^h \psi(z)^2 dz} \quad \#(1)$$

$$Sc = \frac{4\pi \zeta m_e}{\rho d^2} \quad \#(2)$$

It is important to note that due to a damaged bolt, the natural frequency decreased to around 0,71 Hz at the end of the experimental campaign. In that context, an increase of the associated damping ratio, not measured, is also suspected. This will be discussed further in section 5.

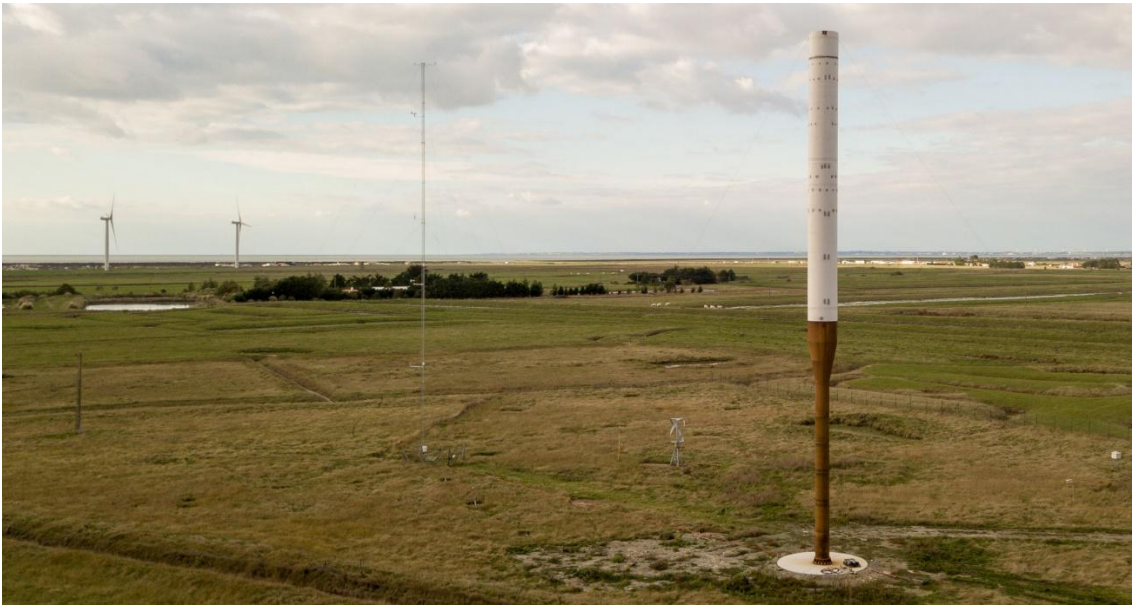

Figure 1 Experimental chimney in the monitored field (the mast with wind anemometers is slightly visible, see figure 2 for details).

## 3.2   Field-test location and instrumentation

The chimney was mounted in a monitored wind field, in Bouin (GPS coordinates 46,975, -1,998), near the Atlantic coast of France. According to the Eurocode [10], this area is in a wind zone category with a 50-years reference wind equal to 26 m/s. The site is surrounded by farmland with sparse gathering of trees and the terrain category is classified as type II [10]. Due to the remote location of the field, and lack of nearby structures, the model chimney has been designed without fearing for loss of human life, animal life or damage to nearby structures.

The chimney was instrumented with two bi-directional accelerometers, one at 20,4 m and one at 35,35 m (near the top) as shown in figure 2. Their measuring range was ±2 g and the acquisition frequency was set to 10 Hz. Wind velocity was measured at several heights using wind anemometers mounted to a 40 m tall truss mast, located 50 m North-West of the chimney (see figure 2). Vane anemometers, measuring speed and direction, were placed at heights of 18 and 35 m and a cup anemometer was located at 10 m. Both Vane and cup anemometers record the wind statistics (mean, standard deviation, maximum of speed and direction for the vane anemometers) over a 10-minute period. An additional sonic anemometer was located at 25 m height. It could measure the unsteady velocity (3 components) at a rate of 5 Hz. While the recording frequency of the sonic anemometer was different from the accelerometers, the recordings were time synchronized. Locations of sensors are sketched in figure 2.

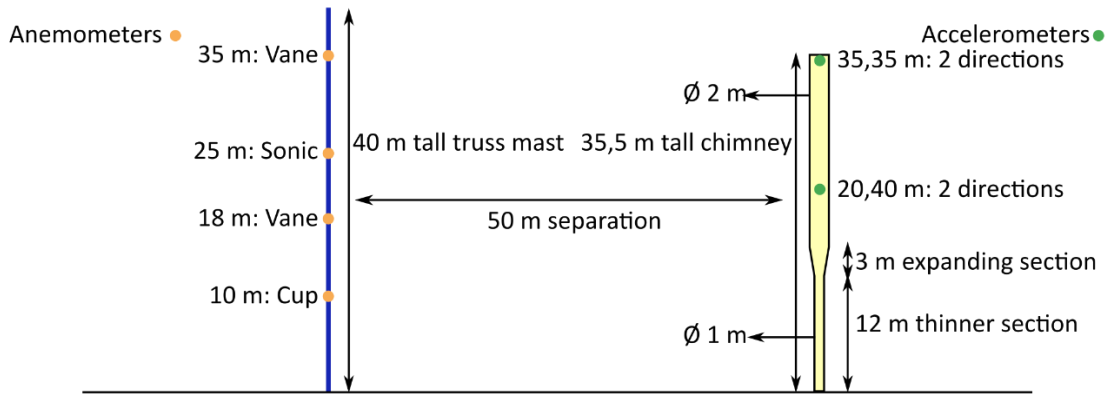

*Figure 2 Sketch of the chimney and anemometer's mast, dimensions and locations of anemometers and accelerometers sensors.*

## 3.3   Data analysis process

Vibration and wind results shown in the present study are based on 1872 ten-minutes records gathered at 35 m over a sequential 13-days period in September 2020. Additional wind data, gathered with all the anemometers distributed along the truss mast, were used to plot the mean and turbulent velocity profiles of the incoming wind.

Each sample of ten-minutes top wind was analyzed by first getting the mean velocity and dominant wind direction. As the accelerometer directions are constant, the displacements are transformed to cross and inline vibrations using the direction of incoming wind. This was used to calculate, using the top bi-directional accelerometer, an associated 10-minute cross-wind acceleration signal. The displacement, *y(t)*, was calculated from the acceleration signal using the inverse Fourier transform of the spectrum $Y(\omega)$ obtained from the Fourier transform of the acceleration $A(\omega)$ and relation (4). A fifth order high-pass Butterworth filter with cutoff frequency of 0,3 Hz has been applied to the acceleration data to eliminate low frequency noise amplified by the Fourier identity [17].

$$A(\omega) = -\omega^2 Y(\omega) \qquad \#(4)$$

The Hilbert transform [8] was used to calculate the response envelopes of the displacement in order to get the mean, maximum and standard deviation of the displacement amplitude over a 10-minute recording. The associated dominant vibration frequency was identified by peak detection on the spectrum *Y(ω)*.

For each ten-minutes sample two cross-wind vibration values were gathered: the maximum amplitude of vibration and the associated dominant frequency along with two wind values: the mean velocity and the dominant direction. Statistical analysis was performed using the 1872 samples in order to plot the probability distribution of the vibration amplitudes and associated dominant frequency as a function of wind velocity or direction. These probability distributions were calculated using a statistical kernel function [31]. The kernel used here is the standard Gaussian kernel used in R and the ggplot2 library (version 4.0.2 and 3.3.2 respectively).

## 4 Wind characterization

### 4.1 Wind speed and directional distribution

The top vane anemometer was used to create the "wind rose" plotted in Figure 3 which shows the distribution of wind speeds and directions. The two most frequent wind directions were north-westerly and north-easterly. Additionally, the most frequently observed speed range was 4 to 6 m/s with speeds above 6 m/s having a long tail and short tail for speeds below 4 m/s. When ignoring the direction, the distribution of all wind speeds resembles a discrete log-normal probability distribution or a negative binomial distribution. The probability of a specific wind direction was found using the relative frequencies and are: S -4,4 %, SW -1,5 %; W -1,9 %; NW -29,5%; N -16,8%; NE-31,4%; E -11,4% and SE -3,1%.

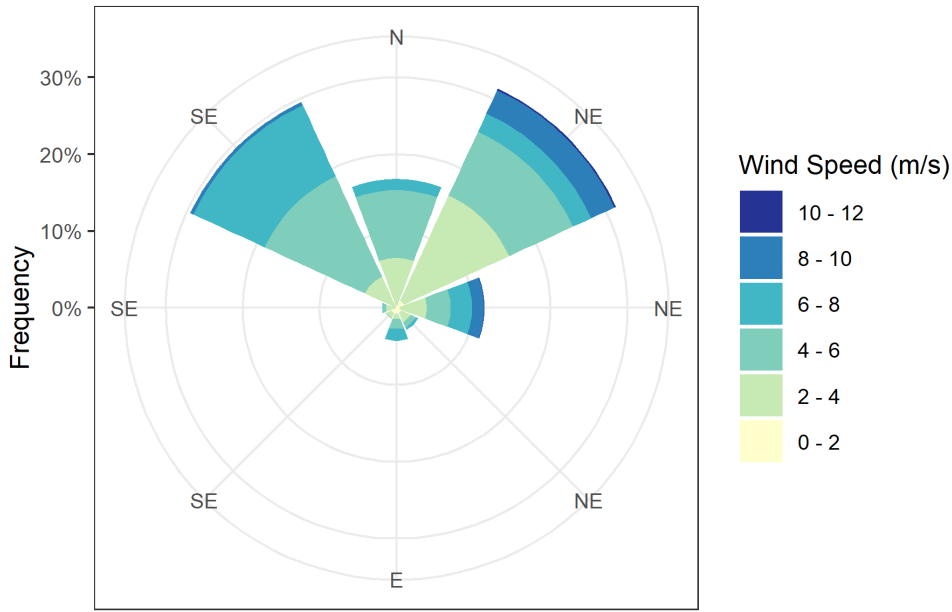

*Figure 3 Frequency of incoming wind velocity (with 8 directional bins) using the 10-mintue mean directions and speeds at 35 m.*

Using the measured wind speeds and $U_{crit}$ = 8,7 m/s, the probability of seeing a speed above $0,8 \cdot U_{crit}$ (Eurocode's recommendations for the onset of VIV), is 21,3 %. As vortex-induced vibrations is only observed in a specific speed range, an upper limit to the investigated wind speed range can be added and is defined as $1,2 \cdot U_{crit}$ which is when one of the Eurocode design methods gives the highest amplitude [10]. With this, 21,3 % of the observed incoming wind speeds can be found in the vortex-induced wind speed range of 7 to 10,5 m/s. The two given percentages are the same as 10,44 m/s was the maximum observed mean wind speed.

### 4.2 Wind speed and turbulence profiles

The mean wind velocity and turbulence intensity evolution with height, for all eight cardinal and ordinal directions are plotted in figure 4, gathering wind data over the anemometer masts, for wind speeds greater than 5 m/s at 35 m height. The turbulence intensity I(z), is defined as the standard deviation of speed at a given height divided by the corresponding mean speed. Eurocode mean wind velocity and turbulence profiles, for terrain category II is also plotted in figure 4.

One can first notice that those mean velocity and turbulence profiles, for which ten-minute wind speed at 35 m remain lower that 12 m/s, strongly depend on the wind direction. While the Eurocode type II mean velocity profile was close to a median profile in comparison with the experiments, the Eurocode type II turbulent profile overestimate the turbulent intensity for all the direction. A direct comparison with Eurocode profiles, which concern reference wind (ten-minutes at 10 meters) of higher mean value (50-years wind), should then be considered with some cautions. Nevertheless, for this low to moderate wind speed campaign one can highlight some relevant information regarding the incoming wind that will be useful for the vibration analysis.

The north-westerly wind which has the highest probability of occurrence has a mean velocity profile close to Eurocode type II model but a turbulence intensity value twice lower with a value slightly less than 10 % at 35 m. The easterly wind which concerns 11,4 % of the observed direction but highlight significant sequences of vortex-induced vibrations of the chimney, was characterized by an important speed gradient with height and a low turbulent intensity less than 2 % at 35 m. This means that wind coming from inland and headed towards the ocean has the strongest shear but the lowest mean turbulence intensity at the heights measured.

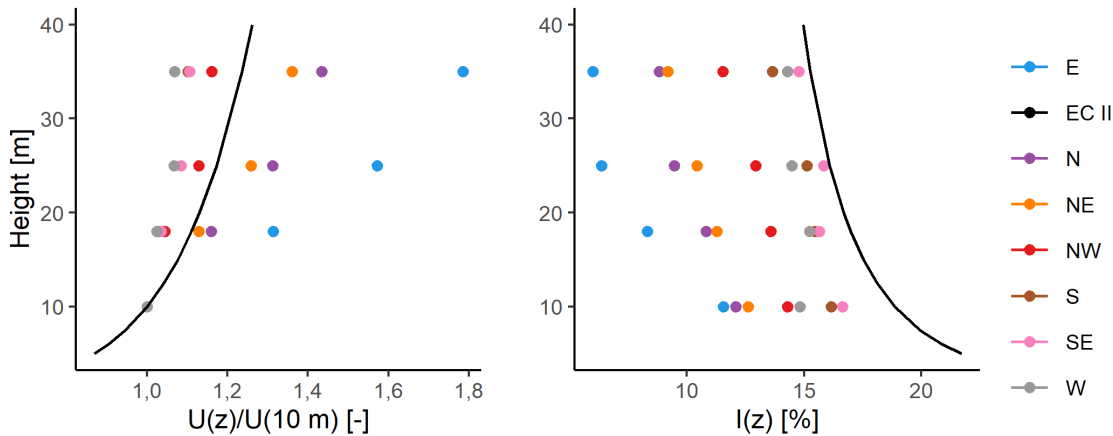

*Figure 4 Mean incoming speed and turbulence profiles (filled dots) compared with the Eurocode profile for terrain category II (solid lines).*

## 5   Cross-wind vibrations of the chimney

Following the data analysis process recalled in section 3.3, a statistical analysis of the chimney's cross-wind vibration was performed. An example of build-up to vortex-induced vibrations and the steadiness of it during lock-in is shown in figure 5. This figure shows the displacement and amplitude envelope for a segment with easterly wind which starts at 4,5 m/s but steadily increased to above 5 m/s according to the sonic data. Statistical distributions of the maximum dimensionless amplitude of vibration (normalized with the top diameter) and associated dominant frequency (normalized with the chimney's natural frequency), are plotted in figures 6 and 7 as a function of wind velocity.

Results are reported using violin plots, mirroring the probability distribution of the data along the y-axis, as a function of discrete wind speed groups (nominal speed value ±0,25 m/s). The exceptions are for 1,25 and 10,25 m/s which groups all speeds below 1,5 and above 10 m/s respectively. In addition to the violin plot, boxplots highlighting the summary statistics (median and quartiles) are shown. A benefit of violin plots over boxplots, is that it shows the distribution of the data. It is particularly relevant for multimodal processes [31] for which the most likely value, associated to the widest point of the violin shape, can differ from the median value. It should also be noted that for the sake of visibility, the mirrored probability distributions shown in figures 6-8 were scaled so that their width were fixed for all speeds. Moreover, the tails of the probability distributions, which contain artifacts of the kernels used, were removed from these plots.

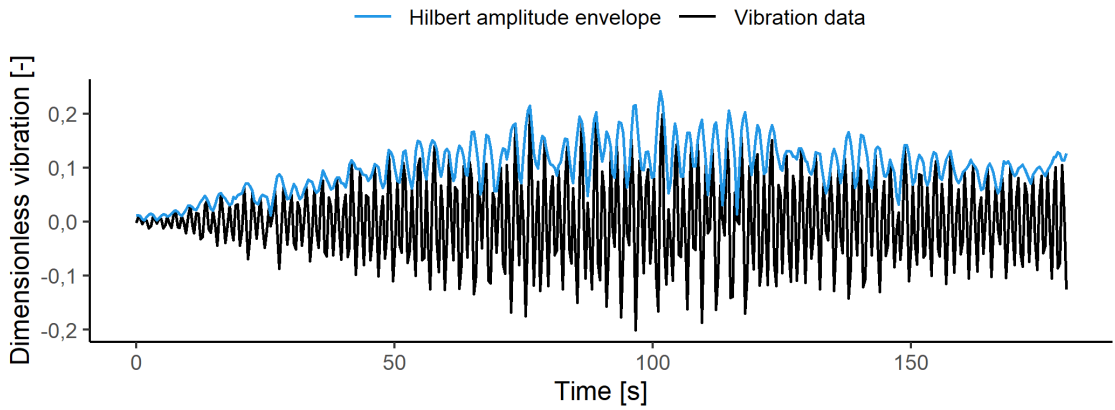

Figure 5 Example of cross-wind displacement signal with build-up to vortex-induced vibrations and the vibration during lock-in at 5 m/s.

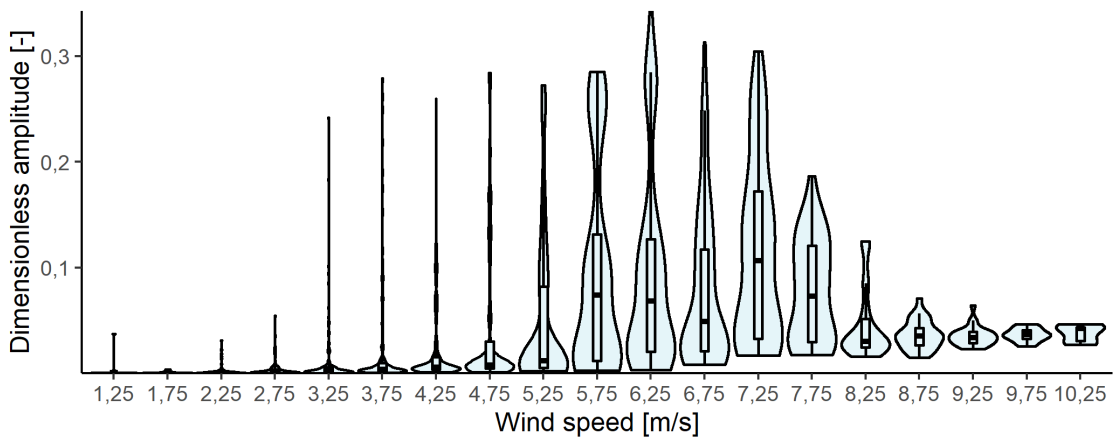

Figure 6 Probability distributions of maximum dimensionless amplitude (normalized with tip diameter) at given speed range. The interior rectangle and lines are boxplots showing summary statistics.

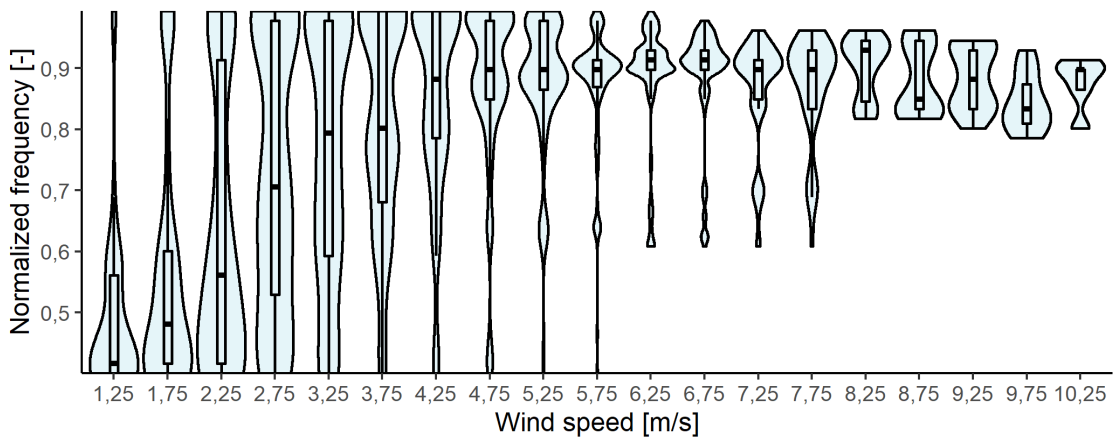

Figure 7 Probability distributions of the dominant frequency of motion (normalized with chimney's first natural frequency) at given speed range. The interior rectangle and lines are boxplots showing summary statistics.

As pointed out in section 4.2, the direction of the incoming wind strongly affects the mean wind velocity and turbulent intensity profiles. One then expects a significant impact of the wind direction on the chimney's cross-vibration. Statistical distributions of the maximum dimensionless amplitude of vibration are then plotted in figure 8 as a function of wind direction. Direction groups used in figure 8 were the cardinal and ordinal directions ±22,5°, with 0°±22,5° defined as northerly wind and 90±22,5° as easterly. Both of the groups were based on the mean values from the vane anemometer at 35 m.

Figure 6 clearly shows that cross-wind vibrations of significant amplitude can be observed for wind speeds between 5 and 8,5 m/s. With a maximum amplitude up to 35% of the diameter (0,7 m) observed at speeds between 6 and 6,5 m/s. This maximum amplitude was close to the one calculated using the Eurocode's method 1 (based on Ruscheweyh's approach [23]) and 35 % lower than the one calculated using the Eurocode's method 2 (based on Vickery and Basu's approach [28]). However, this maximum amplitude was observed at a wind speed 2 m/s lower than the one recommended by Eurocode, suggesting a higher Strouhal number value (closer to 0,25) at high Reynolds number (Re ≈ $8,3 \cdot 10^5$ for U = 6,25 m/s).

At speeds below 5 m/s, there were cases of amplitudes greater than 20 % of the diameter. In most of these cases, the mean speed was slowly reducing from the VIV lock-in speed region over several 10-minute recordings. The vibration amplitude continued to be high and it was possible for high amplitude VIV to continue until mean wind speeds as low as 3,3 m/s. In a few other cases, high amplitudes response could be due to the speed increasing towards the end of the 10-minute recording.

Lower amplitude levels (less than 15 % of diameter) were also observed in the wind speed range 5 to 8,5 m/s. Based on the shape of the violin plot, the lower amplitude vibrations have higher conditional probability than the high amplitude vibrations. The statistical distributions of the maximum amplitude as a function of the wind direction, reported in figure 8, suggest that the lower VIV data are likely to be attributed to north-westerly wind which was the most frequently observed direction and that the sequences of vortex-induced vibrations with the highest amplitude were due to the low turbulent easterly wind.

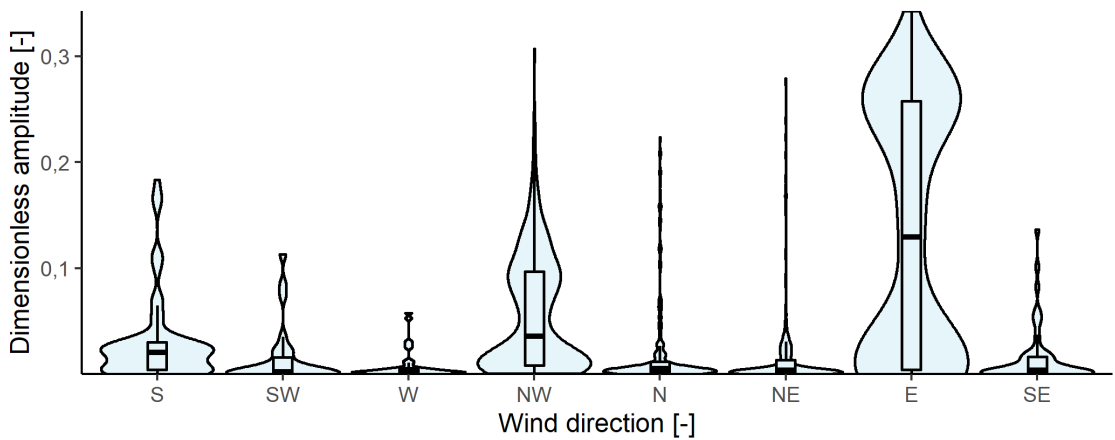

*Figure 8 Probability distributions of maximum dimensionless amplitude (normalized with tip diameter) for different directions ±22,5°. The interior rectangle and lines are boxplots showing summary statistics.*

No vibration amplitudes greater than 7,5 % of the diameter were observed for wind speeds higher than 8,5 m/s. From 8,5 m/s up the 10,25 m/s (which groups all speeds above 10 m/s), one can observe that the probability distributions of the maximum amplitude are more centered with a median value gradually increasing with the velocity. The vibrations can then be due to turbulence-induced vibrations. In this range of wind velocities, the probability distributions of dominant frequency are more surprising. For turbulence-induced vibrations one would expect a dominant frequency close to the first natural frequency of the chimney (and thus a dominant normalized frequency close to 1). Nevertheless, the shape of the violin plot reveals two areas of high probability for the dominant frequency (see for example figure 7, for U = 8,75 m/s), one with normalized frequencies between 0,9 and 1 and the other with normalized frequency between 0,8 and 0,9. As reported in section 3.1, a damaged bolt was observed at the end of the test campaign. A check showed a decrease of 10 % for the chimney's first natural frequency that could explain this peculiar distribution of the dominant frequency in this turbulence-induced vibration regime.

Even if the probability of occurrence of cross-vibrations with significant amplitude was rather low below 5 m/s, it is interesting to focus on the evolution of the probability distribution of the dominant frequency with wind speed before the lock-in (see figure 7). Up to 4,75±0,25 m/s, two distinct frequency groups can be highlighted. One group contains the median frequency which increases almost linearly with the wind speed while the other group concern normalized frequencies between 0,9 and 1 with increasing conditional probability with speed. The first group is clearly related to the vortex shedding signature while the second is due to turbulence-induced vibrations.

Relevant information on response can be found in figure 8 which shows the statistical distributions of maximum amplitude as a function of the wind direction. Easterly (low turbulent) winds were the most favorable to generate high amplitude (>30 % of top diameter) vortex-induced vibrations. Easterly winds also have the highest conditional probability of maximum amplitudes greater than 15 % of the diameter (near 50 % of the maximum amplitudes were above 15 % of the diameter). Vortex-induced vibrations were also observed with the more turbulent north-westerly winds, but with lower amplitude of vibrations. Vibrations up to 18% of the diameter was observed for southerly winds but with a low probability of occurrence. The shape of the violin plot for southerly direction shows a high conditional probability for vibrations lower than 5 % of the top diameter and this could be due to turbulence induced vibration (according to figure 4, the turbulent intensity was high, close to 15 % at 35 m height, for southerly winds).

The conditional probability of wind speeds in the range 5 to 8 m/s was higher for easterly, north-westerly and southerly winds (it was 59,6 % for north-westerly wind, 47,6 % for southwardly and 42,7 % for easterly wind, see figure 3). For northerly and north-easterly wind, on the other hand, it's much more likely to see speeds below 5 m/s and the most likely amplitudes are low. This might be a reason for why the first three mentioned directions have higher conditional probability for VIV of significant amplitude. Easterly wind (towards the ocean) also has the strongest speed gradient with height and the lowest turbulent intensity (less than 2 % at a height of 35 m in comparison to 12 % for the north-westerly wind). While the full impact of shear flow on vortex wake signature and VIV is not well understood, it is well known that VIV is stronger for low turbulent flow in 2D experiments [1, 11, 22, 26, 29]. This is also shown in the present study in the presence of atmospheric boundary layers with different turbulence intensity profiles.

# 6  Conclusion

A custom-made chimney with large top diameter (2 m) and low Scruton number (Sc = 1,82) was erected in a monitored wind field, near the Atlantic coast of France. Details on the field-test platform and methodology, including the structural characteristics of the chimney and the wind "potential", have been presented. Preliminary vibration results, obtained during a sequential 13-days period in September 2020, was presented and discussed. Amplitude and frequency responses were reported in term of probability distributions plotted as a function of both wind speed and direction. As expected two types of cross-wind vibrations were observed, turbulence-induced vibrations and vortex-induced vibrations.

Vortex-induced vibrations of significant amplitude were mostly observed for wind speeds between 5 and 8,5 m/s with maximum amplitude near 6,25 m/s. This "critical" velocity value was lower than expected, suggesting a higher Strouhal number (closer to 0,25) for high Reynolds numbers (Re ≈ 8,3·$10^5$ for U = 6,25 m/s). Several VIV events of low (< 15 % of diameter) to moderate amplitude (> 30 % of diameter) were observed with lower amplitudes being more likely. The results also show that easterly (low turbulent) winds were mainly responsible for the highest amplitudes of vibration (> 30 % of top diameter) while the low amplitude VIV response were mainly due to north-westerly and south winds with higher turbulence intensity.

The goal of this test platform is to gather VIV data at supercritical Reynolds number in real atmospheric wind. These preliminary results will help to forecast specific VIV events on this chimney in order to strengthen the present results and perform additional unsteady pressure measurement, to better understand the 3D vortex signature, loading and VIV response at "super-critical" Reynolds numbers and for different turbulence conditions.

# 7 Acknowledgement

This work is part of a partnership co-funded by Beirens (of the Poujoulat Group), Centre Scientifique et Technique du Bâtiment (CSTB), Centre National d'Etudes Spatiales (CNES) and LadHyX, CNRS-Ecole polytechnique. Special acknowledgement is extended to Aurélien Jeanneton (of Beirens) for designing and constructing the chimney used in the field experiment.